\documentclass[aps,preprint,unsortedaddress,floats]{revtex4}
\usepackage{amsmath,amsfonts,graphicx}

\begin{document}
\bibliographystyle{revtex}
\draft

\textwidth 16cm \textheight 23cm \topmargin -1cm \oddsidemargin
0cm \evensidemargin 0cm

\title{Influence of the Lower Hybrid Drift Instability on the onset of
Magnetic Reconnection}

\author{Paolo Ricci$^{a,b}$}\email{paolo.ricci@polito.it}
\author{J.U. Brackbill$^{c}$} \email{jub@lanl.gov}
\author{W. Daughton $^{c}$}\email{daughton@lanl.gov}
\author{Giovanni Lapenta$^{a,c}$}\email{lapenta@lanl.gov}

\affiliation{a) Istituto Nazionale per la Fisica della Materia
(INFM), Unit\`a del Politecnico di Torino, Corso Duca degli
Abruzzi 24 - 10129 Torino, Italy}

\affiliation{b) Dipartimento di Energetica, Politecnico di Torino,
Torino, Italy}

\affiliation{c) Los Alamos National Laboratory, Los Alamos NM
87545}

\date{\today}

\begin{abstract}

Two-dimensional and three-dimensional kinetic simulation results
reveal the importance of the Lower-Hybrid Drift Instability LHDI
to the onset of magnetic reconnection. Both explicit and implicit
kinetic simulations show that the LHDI heats electrons
anisotropically and increases the peak current density. Linear
theory predicts these modifications can increase the growth rate
of the tearing instability by almost two orders of magnitude and
shift the fastest growing modes to significantly shorter
wavelengths. These predictions are confirmed by nonlinear kinetic
simulations in which the growth and coalescence of small scale
magnetic islands leads to a rapid onset of large scale
reconnection.

\end{abstract}

\maketitle

\section{Introduction}

Understanding the onset of magnetic reconnection remains an open
unsolved question. Reconnection is observed in a great variety of
systems, from special purpose laboratory experiments
\cite{Gekelman1991, Yamada1999, Egedal2003, Furno2003}, to fusion
devices \cite{Taylor1986} and space \cite{Oieroset2001,
Nishida1978, Priest1982} and astrophysical plasmas
\cite{Romanova1992, Blackman1996, Lesch1997}. Our focus is on
reconnection in the magnetosphere due to interactions with the
solar wind. Reconnection develops mainly in current sheets at the
magnetopause \cite{Nishida1978} and in the tail
\cite{Oieroset2001}.

In both situations, the field can be approximated by a Harris
current sheet \cite{Harris1962}. In fact, a reference
configuration for such plasmas is proposed, the so-called GEM
challenge, to bring uniformity, repeatability and ease of
inter-comparison among various models (see Ref. \cite{Birn2001}
and references therein).

The original GEM challenge is initialized with a large magnetic
island (perturbation of $10\%$ amplitude) so that both
magnetohydrodynamic (MHD) modeling and kinetic simulations can
study reconnection, {\it after} reconnection starts.  Thus, the
GEM challenge bypasses reconnection onset and allows  researchers
to study the physics of reconnection in its nonlinear phase. In
our simulations, we consider a GEM equilibrium in which an initial
perturbation is absent so that we can study reconnection onset.

There have been many previous studies of reconnection onset. Among
kinetic studies, which are most relevant to collisionless
magnetospheric plasmas where the current sheet thicknesses are of
the order of the ion gyroradius or ion skin depth, the onset of
reconnection has been long attributed to the tearing instability
\cite{Coppi1966, Drake1976, Galeev1982}.

However, the tearing instability saturates at low levels: For
typical magnetospheric configurations, tearing instability
saturation levels are so low that they yield tiny magnetic islands
with widths of the order of the electron skin depth
\cite{Biskamp1970}. These are far too small to trigger
reconnection by any of the recently investigated, nonlinear
mechanisms for fast reconnection \cite{Biskamp1970}. The GEM
challenge, with its large initial perturbation, produces a large
enough starting island size to decouple electron and ion motion.
This allows fast reconnection through Hall or whistler physics
reconnection \cite{Sonnerup1979, Terasawa1983, Shay1998, Shay2001,
Rogers2003, Ricci2004b}. However, without an initial perturbation,
it is difficult for tearing modes alone to produce sufficiently
large islands, and fast reconnection simply does not occur.
Conceivably, the tiny islands could coalesce into larger islands
until they reach the critical size necessary for fast reconnection
physics. However, this mechanism is quite slow, and the present
paper seeks other mechanisms that occur on much faster time
scales. The tearing instability is also believed to be suppressed
by the presence of even a relatively small vertical magnetic field
(i.e., orthogonal to the current sheet)  \cite{Sergeev1993,
Pellat1991, Pritchett1994, Quest1996, Sitnov2002}. We defer this
issue to a later study.

Our study of the onset of collisionless magnetic reconnection in a
GEM challenge Harris sheet equilibrium \cite{Birn2001} without an
initial perturbation shows that the Lower Hybrid Drift Instability
(LHDI) (see Ref. \cite{Biskamp2000} for a recent review of the
literature) has an important role in the onset of reconnection in
three dimensions.  In two dimensions, tearing instabilities
saturate at low amplitudes, as expected. In three dimensions, the
LHDI modifies the current sheet in ways that cause the tearing
mode to grow more vigorously, as also observed by Sholer {\it et
al.} \cite{Zeiler2003}, and by  Shinohara and Fujimoto
\cite{Shinohara2003}.

Waves at the lower hybrid frequency are observed near reconnection
sites \cite{Carter2002a, Carter2002b, Shinohara1998}, and are
often considered as a source of anomalous resistivity.  In the
earliest model of reconnection presented by Sweet and Parker,
based on resistive MHD, the classical Spitzer resistivity is not
able to explain the reconnection rates observed.
Microinstabilities can enhance the plasma resistivity beyond the
classical values, and thus, the LHDI, which is unstable to a broad
range of wavelengths and frequencies, has been studied extensively
as a source of anomalous resistivity. The fastest growing modes
have wavelengths $k_y \rho_e \approx 1$, $\omega \approx
\Omega_{lh}$, are active in the low-$\beta$ region of the current
sheet and are predominantly electrostatic \cite{Davidson1977}.
Longer wavelength modes ($k_y \sqrt{\rho_i \rho_e} \approx 1$,
$\gamma \approx \Omega_{ci}$) have a large electromagnetic
component and for sufficiently thin sheets can penetrate into the
center of the current sheet \cite{Brackbill1984, Daughton2003}.
However, observations show that the LHDI is confined to the edge
of current sheets, its amplitude seems uncorrelated with the
reconnection electric field, and it saturates at too low amplitude
to explain the enhanced reconnection rate through an anomalous
resistivity \cite{Carter2002a, Carter2002b, Shinohara1998}.

Recently, the effects of the Lower-Hybrid drift instability have
been revisited. It has been observed that in the non-linear phase
of the evolution of the LHDI in a Harris sheet, modifications of
the profiles are induced.

First, the current sheet is thinned and the electron current
profile is peaked \cite{Horiuchi1999, Daughton2002,
LapentaBrack2002, Lapenta2003}. The effect of electron current
sheet peaking is primarily due to electron acceleration and not to
electron density modification \cite{Zeiler2003}. From these
results the possibility emerges that the modifications of the
current profile induced by the LHDI allow the onset of secondary
instabilities and modify the growth rates of instabilities already
present in the initial equilibrium.

Second, previous results based on two-dimensional simulations in
the plane of the LHDI (orthogonal to the magnetic field and to the
plane where  tearing develops) have indeed shown that a
Kelvin-Helmholtz Instability (KHI) develops at much faster rates
than predicted by linear theories based on the initial equilibrium
(see Ref. \cite{Lapenta2003} and references therein).

Furthermore, previous three-dimensional simulations
\cite{Zeiler2003, Lapenta2000, Shinohara2003} have suggested that
the modification of the initial profile caused by the LHDI can
allow the onset of reconnection and enhance the growth of the
tearing instability.

In the present paper we revisit the issue with more detailed
simulations and with more extensive diagnostics, by which we uncover new
physics and confirm previous results.

First, we discover that the nonlinear evolution of the LHDI not
only heats electrons (an effect documented in many previous works,
e.g. Ref. \cite{Brackbill1984}) but preferentially heats electrons
perpendicular to the magnetic field. Anisotropic heating has a
great impact on the onset of reconnection. Previous work
\cite{Coppi1983, Chen1984, Chen1985, Chen1988} suggests that
systems with an electron (or an ion) anisotropy are more unstable
to the tearing instability when the perpendicular temperature
exceeds the parallel temperature. We conduct a number of
simulations and we use the linear Vlasov theory to calculate the
effect of temperature anisotropy on tearing onset.

Second, we confirm that the current sheet is thinned by the LHDI.
The thinning can be explained by nonlinear effects of the LHDI
\cite{Daughton2004}. The thinning has the direct effect of
promoting the growth of the tearing instability. We conduct a
number of simulations and theoretical studies based on linear
theory to estimate the effects of current sheet thinning on the
linear growth rate.

In Sec. II, we describe our physical system  and numerical
approach. In Sec. III, we present our simulation results. In Sec.
IV, we discuss our results and conclusions.

\section{Physical system and the numerical approach}

A Harris current sheet equilibrium is considered
\cite{Harris1962}, with an initial magnetic field given by
\begin{equation}
\mathbf{B}_{0}(z)=B_0 \tanh(z/ L) \mathbf{e}_x
\end{equation}
and a plasma density given by
\begin{equation}
n_0(z)=n_0 \, \hbox{sech} ^2(z/ L)
\end{equation}

In the present paper, similar parameters to the GEM challenge are
used \cite{Birn2001}. In the GEM challenge, the current sheet
thickness is $L =0.5 d_{i}$, the temperature ratio is $T_i/T_e=5$,
and the ion drift velocity in the $y$ direction is
$V_{i0}=1.67V_A$. Rather than the standard mass ratio,
$m_i/m_e=180$ is used. The Alfv\'en velocity, $V_A$, and the ion
inertial length, $d_i=c/\omega_{pi}$, are defined with the density
$n_0$ and the field $B_0$. In contrast to the GEM challenge, no
background density is introduced. This is an important point,
since the presence of a background is strongly stabilizing to the
LHDI.

The standard dimensionless parameters necessary to characterize a
Harris current sheet can thus be summarized as,
${\omega_{pe}}/{\Omega_{ce}}=2.88$,  ${\rho_i}/{L}=1.828$,
${m_i}/{m_e}=180$, and ${T_i}/{T_e}=5$, with $v_{th,s}\equiv
\sqrt{2T_s/m_s}$.

The boundary conditions for the particles and fields are periodic
in the $x$ and $y$ directions.  Conducting boundary conditions are
imposed for the fields at the $z$ boundaries while reflecting
boundary conditions are used for the particles. In order to study
the reconnection onset and in contrast to the GEM challenge
\cite{Birn2001}, the Harris equilibrium is initially unperturbed.

We simulate the dynamics in both the reconnection, $(x,z)$, and
current--aligned, $(y,z)$,  planes. In the $(x,z)$ plane,  the
domain is $L_x \times L_z = 25.6 L \times 12.8 L$, corresponding
to $L_x \times L_z = 12.8 d_i \times 6.4 d_i$. (The GEM challenge
domain is $L_x \times L_z = 51.2 L \times 25.6 L$.) In the $(y,z)$
plane, the domain is $L_y \times L_z = 32L \times 12.8L$. We also
perform a three-dimensional simulation, for which the domain is
$L_x \times L_y \times L_z = 25.6 L \times 16 L \times 12.8 L$.
(This simulation is performed with CELESTE3D only, which is
described below.)

To investigate the evolution of the system, a kinetic linear code
and two nonlinear PIC simulation codes are used. The linear Vlasov
results for the tearing tearing mode are calculated using the
formally exact approach described in \cite{Daughton1999,
Daughton2003}.  This technique employs a normal mode calculation
using a full Vlasov description for both ions and electrons.  The
orbit integrals arising from the linear Vlasov theory are treated
numerically using the exact unperturbed particle orbits and
including the form of the perturbation inside the integral.  Both
electromagnetic and electrostatic contributions to the field
perturbation are retained and resulting system of
integro-differential equations is solved using a finite element
expansion of the eigenfunction \cite{Daughton2003}. The basic
strategy involves a normal mode calculation for perturbations of
the form
\begin{eqnarray}
\label{perturbation}
\hat{\phi} &=& \tilde{\phi}(z) \, \exp(-i \omega t +i k_y y +i k_x x) \;, \\
\;\;\;\;\;\;\;\;\;\;\;\;\;\; \hat{{\bf A}} &=& \tilde{{\bf A}}(z)
\, \exp(-i \omega t +i k_y y +i k_x x) \;, \nonumber
\end{eqnarray}
where the complex functions $\hat{\phi}$, $\hat{\bf A}$ are the
perturbed electrostatic and vector potentials.  For a given Vlasov
equilibrium and for a given choice of wavector ($k_x$,$k_y$), the
code computes the real frequency, growth rate (real and imaginary
part of $\omega$) and the complex eigenfunctions $\tilde{\phi}(z)$
and $\tilde{{\bf A}}(z)$ which describe the mode structure.

The nonlinear dynamics are simulated by two PIC codes, an explicit
simulation code NPIC, and an implicit simulation code CELESTE3D.

The explicit plasma simulation code NPIC is based on a well-known
explicit electromagnetic algorithm \cite{Morse1971, Forslund1985}.
The particle trajectories within NPIC are advanced using the
leapfrog technique, and particle moments are accumulated with area
weighting.  The simulations are run on a parallel computer, using
domain decomposition with calls to the MPI library. (In the
present work, the explicit simulations are run on the Los Alamos
Q-machine using as many as 128 nodes.)

The implicit plasma simulation code, CELESTE3D,  solves the full
set of Maxwell-Vlasov equations using the implicit moment method
\cite{Brackbill1985, Vu1992, Ricci2002a}. Both Maxwell's and
Newton's equations are discretized implicitly in time.  The
implicit simulations are run on a workstation.

The non-linear simulations are performed by the two codes with
very different simulation parameters. In the tearing plane, NPIC
employs a $N_x \times N_z=1280 \times 640$ grid, a time step
$\Omega_{ce} \Delta t=0.03$, and $160 \cdot 10^6$ particles.
CELESTE3D uses a $N_x \times N_z=64 \times 64$ grid, with time
step $\Omega_{ce} \Delta t=0.45$, and a total of $5 \cdot 10^5$
computational particles. In the current aligned plane,  NPIC
employs $N_y \times N_z= 1600 \times 640$ grid, a time step
$\Omega_{ce} \Delta t=0.03$, and $2 \cdot 10^8$ particles , while
CELESTE3D uses a $N_x \times N_z= 128 \times 64$ grid, with time
step $\Omega_{ce} \Delta t=0.7$, and a total of $1 \cdot 10^5$
computational particles. The simulation parameters are summarized
in Table I.

Detailed comparison of the plasma dynamics in NPIC and CELESTE3D
are made in a study of the non-linear phase of magnetic
reconnection in plasmas with different $\beta$ values
\cite{Ricci2004b}. The comparison shows that the physical
mechanisms revealed by the two codes agree, which increases
confidence in their validity. Similar comparisons are made with
other codes for the GEM challenge, which show that results with
CELESTE3D are comparable in detail with those of explicit
simulations \cite{Ricci2002a, Ricci2002b}.  The same kinds of
comparisons are made in this study.

\section{Simulation Results}

To study the onset of reconnection, we simulate the reference
configuration defined above, similar to the GEM challenge
\cite{Birn2001}, but without initial perturbation, without
background plasma, $m_i/m_e=180$, and smaller box size. It is left
to the natural noise of PIC to excite any instability. The result
is strikingly different from published GEM challenge results.  The
tearing mode saturates at low amplitude, and the reconnected flux
is a small fraction of the available. If a third dimension in the
current aligned direction, $y$, is added to the simulation of the
same physical system, the tearing mode grows to much larger
amplitude before saturating and reconnection occurs. Two effects
of the LHDI instability appear to cause a dramatic change in both
the linear and the non-linear phase of the tearing instability in
three-dimensional dynamics, compared with the two-dimensional
dynamics.  They are anisotropic electron heating and current sheet
thinning. Both of these effects enhance the linear growth rate of
the tearing instability and strongly affect and increase the
saturation amplitude. In the following subsections, the results of
the simulations are described in detail.

\subsection{Linear tearing and saturation}

\subsubsection{Two-dimensional simulation}

We consider a two-dimensional simulation in the tearing plane
$(x,z)$, with the parameters defined above, similar to the GEM
challenge. For this configuration, linear theory predicts that the
fastest growing tearing mode has wavelength $k_xL=0.5$ and growth
rate $\gamma=0.176 \Omega_{ci}$. In Fig. 1, the amplitude of
tearing modes with $m_x=1,2,3$, and $4$, corresponding to
$k_xL=0.25,0.5,0.75$, and $1$, are shown as a function of time.
The growth of higher mode numbers is negligible.

At the beginning of the simulation, Fig. 1 shows the fast growth
of the tearing mode with $m_x=2, k_x L=0.5$, in agreement with the
linear theory.  The growth of the mode with $m_x=1$ reveals the
merging of two islands.

The tearing instability saturates at a low level. The half-width
of the island at saturation is $w \approx 0.46L$ at time
$t\Omega_{ci}=83$. By comparison, GEM challenge reconnection
encompasses the whole domain at $t \Omega_{ci} \approx 30$ ($w
\approx 10L$). CELESTE3D and NPIC agree in showing low amplitude
saturation of the tearing mode, but the coarser grid spacing used
in CELESTE3D does not allow an accurate estimate of the island
half width. It should be noted that the growth of tearing modes in
NPIC with $m_x=3,4$ at $t \Omega_{ci} \approx 80$ is possibly due
to the numerical heating always present in explicit PIC
simulations like NPIC, which can in principle affect the results
of long time scale simulations. In any case, this slow growth at
late time in the two-dimensional tearing simulation is several
orders of magnitude slower than the growth and non-linear
development of the LHDI in three-dimension that is the subject of
the present work.

The saturation of tearing has been studied theoretically in high
and low-$\beta$ plasmas (e.g., see Refs. \cite{Biskamp1970,
Drake1977, Kuznetsova1990}).  In high-$\beta$ plasmas, like the
ones considered in the GEM challenge, anisotropic heating of
electrons causes saturation of tearing \cite{Biskamp1970}. During
the growth of the tearing instability,
$(T_{eyy}+T_{ezz})/2T_{exx}$ (defined as $T_{e\perp}/T_{e ||}$) is
reduced below 1, and $T_{e, \perp}/T_{e||}<1$ has a strongly
stabilizing effect on tearing. In fact, in the two-dimensional
simulations of tearing at time $t \Omega_{ci}=80$, NPIC gives
$T_{e\perp}/T_{e ||}=0.87$ and CELESTE3D gives $T_{e \perp}/T_{e
||}=0.83$. For $T_{e \perp}/T_{e ||}=0.87$ the linear code
predicts the mode $m_x=2$ ($k_xL=0.5$) to be stable and the mode
$m_x=1$ ($k_xL=0.25$) to be weakly unstable ($\gamma=0.04
\Omega_{ci}$). However, this growth rate is based on a
bi-Maxwellian electron distribution. The simulated velocity
distribution is more complex and might yield a different growth
rate.

It should be remarked that in simulations performed in a more
extended domain, that small islands can coalesce repeatedly into
larger islands \cite{Karimabadi2004}. If the system is big enough,
the final merged island can grow enough to allow the fast
reconnection physics shown by the GEM challenge. However, this
coalescence process occurs on a long time scale compared to the
time scales typical of the three-dimensional instabilities that
are the subject of the present manuscript.

\subsubsection{Three-dimensional simulation}

CELESTE3D is used to perform a three-dimensional simulation of the
Harris current sheet described in Sect. II, in a domain with $L_x
\times L_y \times L_z = 25.6 L \times 16 L \times 12.8 L$.
Tearing, current aligned (i.e., LHDI and KHI), and oblique
instabilities are all involved in the evolution of the system.

The results of three-dimensional simulations performed by
CELESTE3D have been directly compared with satellite observations
\cite{Ricci2004a}. A close agreement has been observed between
simulation and satellite observations, concerning current sheet
kinking, current bifurcation, and reconnecting modes.

In Fig. 2, the amplitude of the Fourier component of the $B_x$
magnetic field is plotted as a function of the mode number $m_x$
(aligned with the magnetic field) and $m_y$ (aligned with the
current) at sequential times. The plots average over $z$ and Fast
Fourier Transform (FFT) in $(x,y)$. The average over $z$
suppresses odd parity modes in $B_x$  and thus only even parity
modes, such as KHI and LHDI, appear in Fig. 2.  Since $B_x$ and
$B_z$ have opposite parity,  the even modes, like the tearing
modes (not shown in Fig. 2) are picked up by $z$-averages of
Fourier modes of $B_z$.

Fig. 2 shows the development of the initial noise, which excites
mostly low mode number instabilities $m_x$ and $m_y$ at $t
\Omega_{ci}=0.2$. The presence of the electrostatic LHDI is
reflected at time $t \Omega_{ci}=2.5$ through a significant
amplitude of the mode number $m_x=0, m_y=12-14$, which corresponds
to $k_y \rho_e \approx 0.2$. The electrostatic LHDI, studies of
which are summarized in Ref. \cite{Biskamp2000},  is dominated by
an electrostatic component on the edge of the current sheet. The
characteristic wavelength is of order $k_y \rho_e \approx 1$, and
frequency $\omega \approx \Omega_{lh}$ \cite{Davidson1977}. The
electrostatic LHDI instability causes a velocity shear that
enhances the growth rate of the electromagnetic LHDI and causes a
KHI to grow \cite{LapentaBrack2002, Lapenta2003}.  Later, at
$t\Omega_{ci}=5$, the maximum amplitude modes are $m_x=0$ and
$m_y=5-6$, which corresponds to $k_y \sqrt{\rho_i \rho_e} \approx
0.5$).  These longer wavelength modes are the electromagnetic
LHDI, which grows at center of the current sheet and has
wavelength $k_y \sqrt{\rho_i \rho_e} \approx 1$ and growth rate
$\gamma \approx \Omega_{ci}$ \cite{Daughton2003}. Velocity shear
created by the LHDI triggers a KHI at a still later time, so that
by $t\Omega_{ci}=20$ the current sheet is dominated by a single,
domain-sized kink with mode numbers $m_x=0$ and $m_y=1$ ($k_yL
\approx 0.4$).

Figure 3 shows the amplitude of the Fourier modes of the $B_z$
field as a function of $m_x$ and $m_y$ at different times. Again,
averaging over $z$ eliminates the odd parity modes and leaves only
even parity modes.  The modes shown in Fig. 3 are complementary to
the those in Fig. 2. The $B_z$ Fourier components are signatures
of the tearing instability. The tearing instability starts to grow
significantly at $t \Omega_{ci}=5$, with mode number $m_x=4$ and
$m_y=0$ ($k_xL=1$, a higher frequency than the expected from the
linear theory). Subsequently, the 4 magnetic islands merge into 2
islands at $t\Omega_{ci}=10$, and then into a 1 that encompasses
the whole domain by $t \Omega_{ci} \approx 20$. This is comparable
to the reconnection time shown by the GEM challenge simulation in
two dimensions with a large initial perturbation \cite{Birn2001}.
No significant growth of oblique tearing modes is observed during
the simulation.

Two main conclusions can be drawn from the three-dimensional
simulation. First, in contrast to simulations in two-dimensions,
the tearing mode does not saturate at a low amplitude in three
dimensions. Instead,  it encompasses the whole domain in a time
comparable to the GEM challenge simulation with a large initial
perturbation \cite{Birn2001}. Second, no significant growth of
oblique modes is seen in the simulation.

Thus, it can be argued that the current sheet dynamics depends
only on some nonlinear interaction between the tearing
instability, which develops in the $(x,z)$ plane, and
current aligned instabilities, which grow in the $(y,z)$ plane. As
the time scale of instabilities in the $(y,z)$ plane is an order
of magnitude faster than the tearing instability, it follows that
the fundamental dynamics of a three-dimensional current sheet can
be understood more conveniently by performing simulations in the
$(x,z)$ and $(y,z)$ plane, and analyzing the effects of the very
rapid current aligned instability on the more slowly growing
tearing mode.

\subsection{Current sheet modifications by the LHDI}

In simulations performed in the current aligned, $(y,z)$, plane,
two modifications of the initial equilibrium appear to enhance the
tearing growth rate. Both are non-linear consequences of the LHDI.

The LHDI causes anisotropic heating of the electrons. In Fig. 4,
the $y$-averaged electron temperature ratio, $T_{e\perp}/T_{e
||}$, is plotted as a function of $z$ at $t \Omega_{ci}=0$  and at
$t \Omega_{ci}=5$. In both the CELESTE3D and NPIC simulations, an
electron temperature anisotropy develops by  $t \Omega_{ci}=5$
both at the center and on the flanks of the current sheet. The
anisotropy ranges between 2.3 in NPIC and 1.7 in CELESTE3D. The
ratio between the ion perpendicular and parallel temperature,
$T_{i \perp}/T_{i ||}$, at $t \Omega_{ci}=0$ and at $t
\Omega_{ci}=5$, is also shown in Fig. 4. According to both PIC
codes, ion anisotropic heating is smaller than electron heating,
and it is located mostly on the flanks of the current sheet.  As
shown in Fig. 5,  the electron distribution function is not
gyrotropic. In particular, the peak electron perpendicular
temperature in the $z$ direction (cross sheet direction) is about
25\% higher than the temperature in the $y$ direction (current
aligned direction). Moreover, the heating in the $z$ direction is
more focused near the center of the current sheet than the heating
in the $y$ direction.

The LHDI also alters the current profile. It thins the current
sheet and increases the value of the peak current density. The
alteration is primarily due to electron acceleration: the ion and
the electron density modifications are small \cite{Horiuchi1999,
Lapenta2000, Zeiler2003, Shinohara2003}. In Fig. 6, the
$y$-averaged current sheet profile is shown at $t=0$ and at $t
\Omega_{ci}=4$ as a function of $z$. The current profiles obtained
from both PIC codes agree well and show that the peak current is
enhanced by about 40\% over the initial current profile.

The LHDI also creates velocity shear. This latter consequence has
been the subject of a number of papers (see Ref.
\cite{Lapenta2003} and references therein). Its main effect is to
promote the growth of a KHI, but the three-dimensional simulation
shows that the KHI develops after the onset of reconnection and
thus appears to have little effect on reconnection onset.

\subsection{Effects of the LHDI on tearing}

Both anisotropic heating and current peaking enhance
the linear growth rate of the tearing instability and change
dramatically its non-linear evolution. The effects of the two
mechanisms on tearing are considered independently in the
following subsections.

\subsubsection{Electron anisotropy}

Figure 7 shows the value of the tearing growth rate $\gamma$ as a
function of $k_x$ for $T_{e \perp}/T_{e ||}=0.9, 1, 1.5, 2$, and
same GEM challenge parameters (in particular $T_i/T_{e \perp}=
5$). The growth rate of the tearing instability increases appreciably
with the ratio $T_{e \perp}/T_{e ||}$: for $T_{e \perp}/T_{e
||}=2$ the maximum growth rate, $\gamma \approx 3.8 \Omega_{ci}$,
is more than one order of magnitude larger than the maximum growth
rate for $T_{e \perp}/T_{e ||}=1$ ($\gamma \approx 0.18
\Omega_{ci}$). Moreover, the tearing mode with maximum growth rate
shifts to shorter wavelengths with increasing anisotropy: in the case $T_{e \perp}/T_{e
||}=2$ the maximum occurs for $k_xL=2.25$, considerably larger
that the typical  $k_xL=0.5$ seen for isotropic electrons. It
should be remarked that electron heating due to LHDI does not lead
to a simple electron distribution function (see Fig. 5), but
assuming a bi-Maxwellian distribution function, allows us to
estimate the growth rate of the tearing instability.

The tearing eigenmodes for $T_{e \perp}/T_{e ||}=2$ and $T_{e
\perp}/T_{e ||}=1$ are compared in Fig. 8. The most remarkable
difference is in the structure of the  potential $\phi$, which is
larger in the presence of anisotropic electrons and with opposite
sign. Moreover, for anisotropic electrons, the tearing eigenmode
involves a narrower region near the center of the current sheet
and the $A_z$ component is more important.

These results are consistent with the enhancement of the tearing
instability growth rates predicted in previous work
\cite{Forslund1968, Coppi1983, Chen1984, Chen1985, Ambrosiano1986,
Shi1987}, but our results are obtained by integration over exact
orbits, similarly to Ref. \cite{Burkhart1989}.  As an aside, it
should be remarked that our linear analysis predicts that ion
anisotropy plays only a minor role in thin current sheets ($\rho_i
\sim L$) in agreement with earlier work by Burkhart and Chen
\cite{Burkhart1989}.

The nonlinear evolution of the tearing mode changes dramatically
when the electron temperature is anisotropic. An initially
favorable temperature anisotropy neutralizes the effect of the
unfavorable anisotropy created by the growth of tearing, which
would otherwise cause nonlinear stabilization.  The evolution of
the current sheet is followed in two-dimensional PIC simulations
in the $(x,z)$ plane, starting with an anisotropic electron
temperature. The GEM challenge parameters are used with
$T_{i}/T_{e ||}=5$ on the same domain $L_x \times L_z =25.6 L
\times 12.8 L$, but with various electron temperature
anisotropies,   $T_{e \perp}/T_{e ||}=1, 1.5, 2$.   In Fig. 9, the
evolution of the tearing instabilities with mode numbers
$m_x=1-11$ (corresponding to $k_xL=0.25-3$) are shown as a
function of time. In comparison with the case for initially
isotropic electrons (see Fig. 1), high mode numbers grow, forming
a number of small tiny islands. Subsequently, these small islands
merge to form a single large
island (the mode number $m_x=1$ dominates),  which encompass the
whole domain by $t \Omega_{ci} \approx 20$.

In Fig. 10, the reconnected flux is plotted as a function of time
for the three different anisotropies, $T_{e \perp}/T_{e ||}=1,
1.5, 2$, for both NPIC and CELESTE3D simulations. With
$T_{e||}/T_{e \perp}=1$, tearing saturates at a low level. For
anisotropic distributions, $T_{e||}/T_{e \perp}=1.5, 2$,
reconnection involves the whole domain and the growth of the
tearing instability saturates when all the available magnetic flux
is reconnected, at a level very similar to the GEM challenge. The
fast reconnection phase is delayed longer for $T_{e
\perp}/T_{e||}=1.5$ than for $T_{e \perp}/T_{e||}=2$.

Figure 11 shows a plot of the temperature anisotropy as a function
of time for the three simulations with different initial
anisotropies ($T_{e \perp}/T_{e ||}=1, 1.5, 2$). The temperature
ratio $(T_{eyy}+T_{ezz})/2T_{exx}$ that corresponds to $T_{e
\perp}/T_{e ||}$ at $t=0$ is plotted. For an initially isotropic
distribution, $T_{e \perp}/T_{e||}=1$, the ratio decreases as a
function of time and stabilizes the tearing instability. For
initially anisotropic distributions, the ratio $T_{e
\perp}/T_{e||}$ decreases rapidly to a minimum value, $T_{e
\perp}/T_{e||} \approx 0.8$, and then increases with the onset of
fast reconnection.

\subsubsection{Current thinning and peaking}

The thinning of the current sheet and consequent increase in
maximum current density also enhances the linear growth rate of
the tearing instability and changes its non-linear evolution. This
is in agreement with the results of other recent simulation
studies \cite{Zeiler2003, Shinohara2003}.

Figure 12 shows the tearing instability linear growth rate,
$\gamma$, for different values of the current sheet thickness
($\rho_i/L=0.914, 1.828, 3.656$). The growth rate is higher for
thinner current sheets: $\gamma= 0.176 \Omega_{ci}$ for
$\rho_i/L=1.828$ (GEM challenge thickness) compared with $\gamma=
0.632 \Omega_{ci}$ for $\rho_i/L=3.656$. The maximum growth rate
is located at $k_xL=0.5$ in all cases.

The non-linear evolution of the tearing instability is simulated
with both NPIC and CELESTE3D. For sufficiently thin current
sheets, reconnection is not blocked by non-linear saturation of
the tearing mode at low levels and encompasses the whole domain.
In Fig. 13, we show the evolution of the current sheet by plotting
the reconnected flux, $\Delta \Psi$, for the Harris sheet with
half the current sheet thickness as the GEM challenge ($\rho_i/L=
3.656$) from the simulations. A domain size $L_x \times L_z= 12.8
d_i \times 6.4 d_i$ (corresponding to $L_x \times L_z = 51.2L
\times 25.6 L$) is considered. Unlike the standard GEM challenge
case without perturbation, reconnection involves the whole domain
by $t \Omega_{ci} \approx 20$ for NPIC and by $t \Omega_{ci}
\approx 13$ for CELESTE3D. It should be remarked that the fast
reconnection phase is very similar in the two codes. The delayed
start of the fast reconnection phase in NPIC is probably because
more particles are used in NPIC than in CELESTE3D, which reduces
the noisiness of the initial conditions \cite{Ricci2004b}.

\section{Conclusion and discussion}

Reconnection onset is studied using results from two codes
employing very different algorithms: NPIC is a massively parallel
explicit code and CELESTE3D is an implicit-moment method PIC code.
The results from NPIC and CELESTE3D complement and confirm each
other. This degree of cross-checking  between codes, for which the
GEM challenge is designed, is unusual. Because of their
differences in resolution, the agreement between explicit and
implicit results in detail suggests that the important physical
length scales involved in the onset of reconnection are comparable
to or greater than the electron scales.

The simulations have pointed out the important role of the LHDI in
reconnection onset. Our results confirm that the tearing
instability saturates at a low level if no initial perturbation is
added to the Harris equilibrium. Linear growth of the tearing
instability is limited by increase in the parallel electron
temperature, such that an anisotropy develops  with $T_{e \perp}
/T_{e ||} < 1 $ that strongly stabilizes the tearing mode. In
three-dimensions, the tearing instability does not saturate at
small amplitudes. Because the mode spectrum reveals no significant
oblique mode growth,  the current sheet dynamics can be analyzed
by studying the interrelationship between the current aligned
instability that develops in the $(y,z)$ plane and the dynamics of
the tearing modes in the $(x,z)$ plane. The analysis shows that
the LHDI strongly modifies the linear and non-linear evolution of
the tearing instability. The LHDI causes a favorable electron
temperature anisotropy, $T_{e \perp} /T_{e ||}
> 1 $, and thins and peaks the current sheet. Both effects seem
effective in enhancing the
linear growth rate of the tearing instability. With a favorable
anisotropy and a thin enough current sheet, the tearing
instability grows large enough to decouple electrons and ions so that reconnection
can encompass the whole domain.

In the current sheet considered there is no background plasma,
the guide field is not present and no normal component of the
magnetic field is introduced. The influence of these on
reconnection onset needs further investigation.

Both laboratory and satellite observations \cite{Carter2002a,
Carter2002b, Shinohara1998} have pointed out the presence of the
LHDI near reconnection sites, but they have not identified a
connection between this instability and reconnection through
a contributions to anomalous resistivity.

Satellites have measured electron anisotropy during magnetic
substorms. Shinohara {\it et al.} \cite{Shinohara1998} observe
electron anisotropy ($T_{e \perp} / T_{e ||}<1$) at the substorm
onset. A more comprehensive study of the electron anisotropy
during reconnection has been performed by Birn {\it et al.}
\cite{Birn1997a}. Through one-year average data of satellite
measurements, it is shown that an electron anistropy ($T_{e \perp}
/ T_{e ||}>1$) precedes substorm onset. At onset, $T_{e \perp} /
T_{e ||}<1$ is observed.  After onset, $T_{e \perp} / T_{e ||}$
grows again. This behavior recalls closely what is observed in
tearing simulations (see Fig. 11). Birn {\it et al.}
\cite{Birn1997a} also show that an ion temperature anisotropy is
less relevant than an electron temperature anisotropy, and this
also agrees with the simulations presented here.

It is not obvious how to verify current sheet thinning from
observations. Our simulation starts with a plain Harris sheet from
which the LHDI develops and thins the current sheet. In reality,
on the other hand, the LHDI is always present in current sheets
with an amplitude that corresponds to the saturation level for
that particular current sheet. In particular, MRX results
\cite{Carter2002a, Carter2002b} show a thinning of the current
sheet, and a consequent enhancement of the amplitude of the LHDI
instability. It is hard to say if the enhancement of the LHDI is a
cause or a consequence of the thinning process.

\section*{ACKNOWLEDGMENTS}

The authors gratefully acknowledge useful discussions with J.
Chen, B. Coppi, S. Hsu, J. Huba, B. Rogers, and I. Shinohara. This
research is supported by the LDRD program at the Los Alamos
National Laboratory, by the United States Department of Energy,
under Contract No. W-7405-ENG-36 and by NASA, under the "Sun Earth
Connection Theory Program".

\newpage

\begin{itemize}

\item Fig. 1: The amplitude of the tearing mode are shown as a
function of time for the mode numbers $m_x=1$ (dashed), $m_x=2$
(dash-dotted), $m_x=3$ (dotted), and $m_x=4$ (solid)
(corresponding to $k_xL=0.25$, $k_xL=0.5$, $k_xL=0.75$, and
$k_xL=1$, respectively), during the two-dimensional simulation in
the tearing, $(x,z)$, plane. The results are from an NPIC
simulation.

\item
Fig. 2: The amplitude of the Fourier modes of the $B_x$
component of the magnetic field is shown as a function of the mode
numbers $m_x$ and $m_y$ at different times: $t \Omega_{ci}=0.2$
(a), $t \Omega_{ci}=2.5$ (b), $t \Omega_{ci}=5$ (c), $t
\Omega_{ci}=10$ (d), $t \Omega_{ci}=15$ (e), and $t
\Omega_{ci}=20$ (f). The three-dimensional simulation performed by
CELESTE3D is considered.

\item Fig. 3: The amplitude of the Fourier modes of the $B_z$
component of the magnetic field is shown as a function of the mode
numbers $m_x$ and $m_y$ at different times: $t \Omega_{ci}=0.2$
(a), $t \Omega_{ci}=2.5$ (b), $t \Omega_{ci}=5$ (c), $t
\Omega_{ci}=10$ (d), $t \Omega_{ci}=15$ (e), and $t
\Omega_{ci}=20$ (f). The three-dimensional simulation is performed by
CELESTE3D.

\item Fig. 4: The enhancement of the electron temperature ratio
$T_{e \perp }/ T_{e ||}$ (a,c), and of the ion temperature ratio
$T_{i \perp }/ T_{i ||}$ (b,d) due to LHDI is shown from
simulations in the ($y,z$) plane. The temperature ratio is shown
at time $t=0$ (solid line, isotropic distribution) and $t
\Omega_{ci}=5$ (dashed line). The results are from NPIC (a,b) and
from CELESTE3D (c,d).

\item Fig. 5: The electron pressure ratios $P_{ezz}/P_{exx}$ (a)
and $P_{eyy}/P_{exx}$ (b) averaged along $y$ are shown at time $t
\Omega_{ci}=5$. The results are from an NPIC simulation in the
$(y,z)$ plane.

\item Fig. 6: The peaking of the current density $J_y$ is shown
from the simulation in the $(y,z)$ plane. The dotted line
represents the current profile at $t=0$ averaged along $y$, the
profiles at $t \Omega_{ci}=4$ is shown by the solid line (NPIC
simulation) and dashed line (CELESTE3D simulation). The current is
normalized in order that the maximum is equal to 1 at $t=0$.

\item Fig. 7: The growth rate $\gamma / \Omega_{ci}$ of the
tearing mode is plotted as a function of $k_x L$ for $T_{e \perp
}/ T_{e ||}=0.9$ (dotted), $T_{e \perp }/ T_{e ||}=1$ (solid),
$T_{e \perp }/ T_{e ||}=1.5$ (dashed), and $T_{e \perp }/ T_{e
||}=2$ (dash-dotted). The other plasma parameters are the same as
the parameters described in Sect. II (in particular, $T_{i}/T_{e
\perp}= 5$).

\item Fig. 8: An eigenmode solution for the fastest tearing
instability is shown.  The real (solid) and imaginary (dotted)
parts are plotted for $\phi$ (a,e), $A_x$ (b,f), $A_y$ (c,g), and
$A_z$ (d,h) with $T_{e \perp }/ T_{e ||}=1$, $k_x L=0.5$, and
$\gamma=0.176 \Omega_{ci}$ (a,b,c,d) , and $T_{e \perp }/ T_{e
||}=2$, $k_xL=2.25$, and $\gamma=3.782 \Omega_{ci}$(e,f,g,h).

\item Fig. 9: The amplitude of the tearing mode is shown as a
function of time for the mode numbers $m_x=1$ and $k_xL=0.25$
(solid blue), $m_x=2$ and $k_xL=0.5$ (solid green), $m_x=3$ and
$k_xL=0.75$, (solid red), $m_x=4$ and $k_xL=1$, (solid cyan),
$m_x=5$ and $k_xL=1.25$, (solid magenta), $m_x=6$ and $k_xL=1.5$,
(solid black), $m_x=7$ and $k_xL=1.75$ (dashed blue), $m_x=8$ and
$k_xL=2$ (dashed green), $m_x=9$ and $k_xL=2.25$, (dashed red),
$m_x=2.5$ and $k_xL=10$, (dashed cyan), $m_x=11$ and $k_xL=2.75$,
(dashed magenta), and $m_x=12$ and $k_xL=3$, (dashed black),
during the two-dimensional simulation in the tearing plane with
plasma parameters described in Sect. II, but $T_{e \perp}/ T_{e
||}=2$ and $T_i/T_{e \perp}=2.5$. The results are from an NPIC
simulation.

\item Fig. 10: The reconnected flux, $\Delta \Psi$, is shown as a
function of time, for simulations with the parameters described in
Sect. II (solid line), parameters described in Sect. II but $T_{e
\perp}/ T_{e ||}=1.5$ (dotted), and $T_i/ T_{e ||}=5$ (dashed),
and plasma parameters described in Sect. II, but $T_{e \perp}/
T_{e ||}=2$ and $T_i/ T_{e ||}=5$. Both NPIC results (a) and
CELESTE3D results (b) are shown.

\item Fig. 11: The electron temperature ratio $(T_{e yy}+T_{e
zz})/ 2 T_{e xx}$ (corresponding to $T_{e \perp}/ T_{e ||}$ at
$t=0$) is plotted as a function of time, for simulations with the
parameters described in Sect. II (solid line), plasma parameters
described in Sect. II, but $T_{e \perp}/ T_{e ||}=1.5$ (dotted),
and $T_i/ T_{e ||}=5$ (dashed), and parameters described in Sect.
II but $T_{e \perp}/ T_{e ||}=2$ and $T_i/ T_{e ||}=5$. The
results are from NPIC.

\item Fig. 12: The growth rate $\gamma / \Omega_{ci}$ of the
tearing mode as a function of $k_x L$ is plotted for plasma
parameters described in Sect. II and different values of
$\rho_i/L$.

\item Fig. 13: The reconnected flux as a function of time is
plotted for the simulation in the $(x,z)$ plane for a plasma with
the plasma parameters described in Sect. II, but $\rho_i/L=3.656$.
NPIC (solid) and CELESTE3d (dashed) results are plotted.

\end{itemize}

\newpage

\bigskip
\textbf{Table I.} Simulation parameters

\bigskip

\begin{tabular}
[c]{lccc} \hline\hline
& grid & particles & $\Delta t \Omega_{ce}$ \\
\hline
$(x,z)$ plane (NPIC) & $1280 \times 640$ & $16 \cdot 10^7$ & 0.03 \\
$(x,z)$ plane (CELESTE3D) & $64 \times 64$  & $5 \cdot 10^5$ & 0.45\\
$(y,z)$ plane (NPIC) & $1600 \times 640$ & $2 \cdot 10^8$ & 0.03\\
$(y,z)$ plane (CELESTE3D) & $128 \times 64 $ & $1 \cdot 10^5$ & 0.7\\
\hline\hline
\end{tabular}

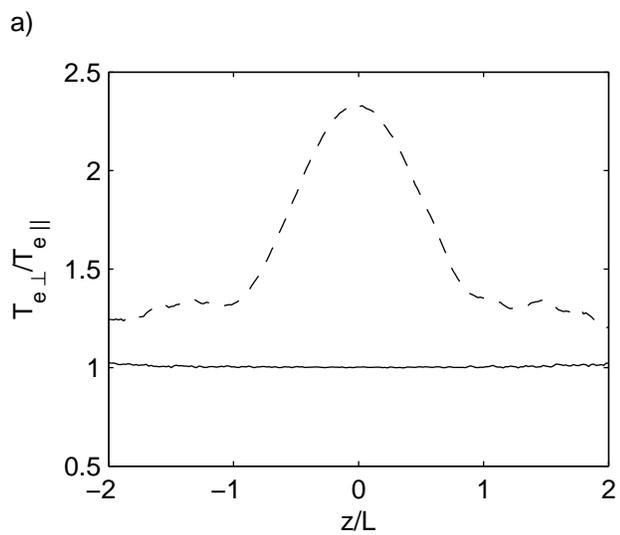
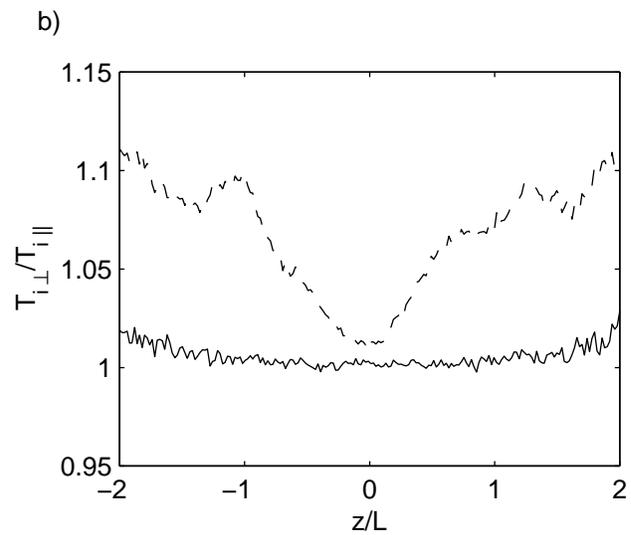
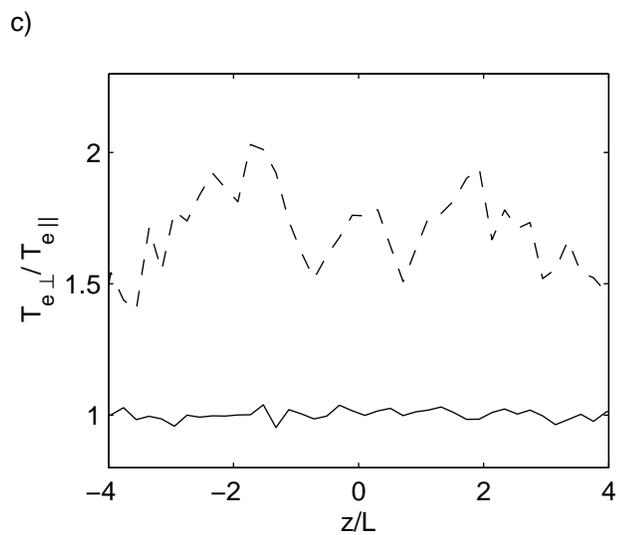
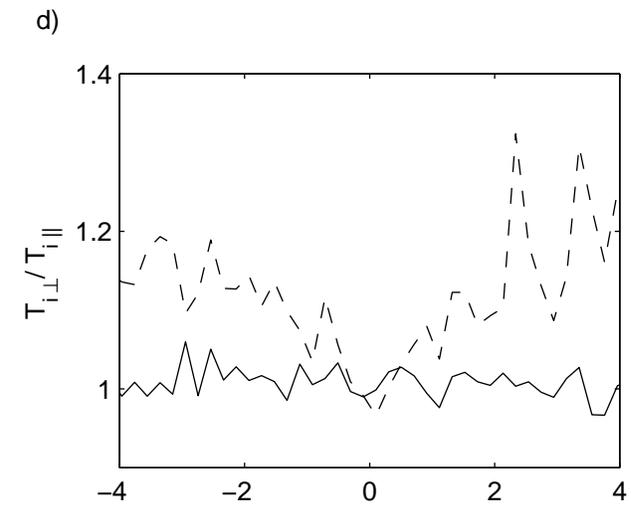
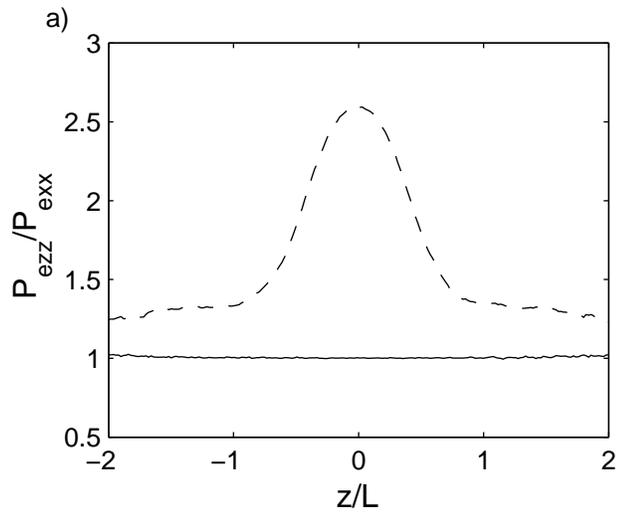
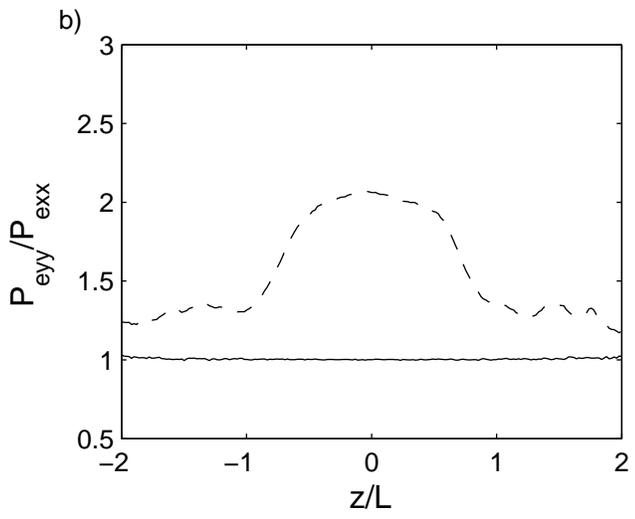
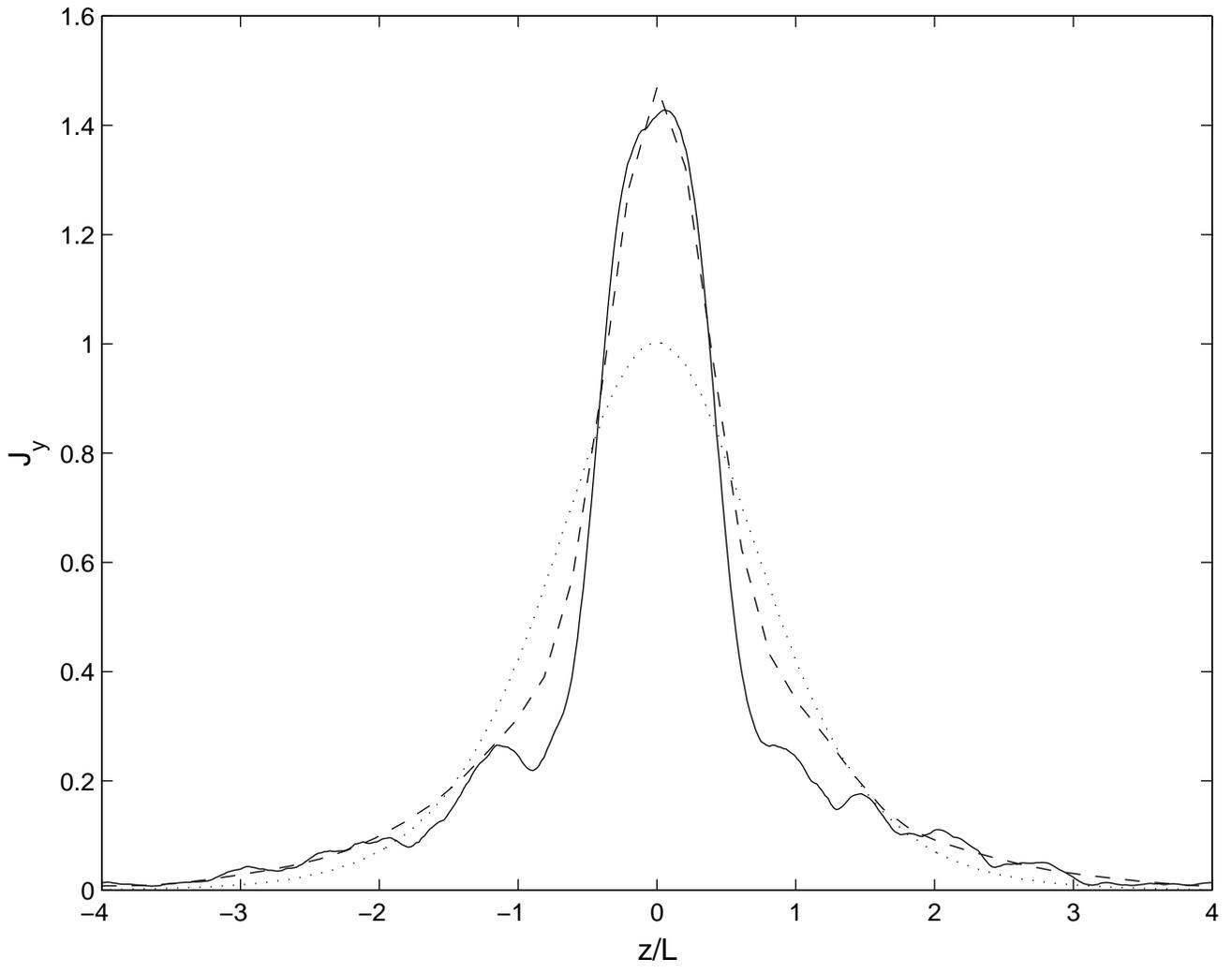
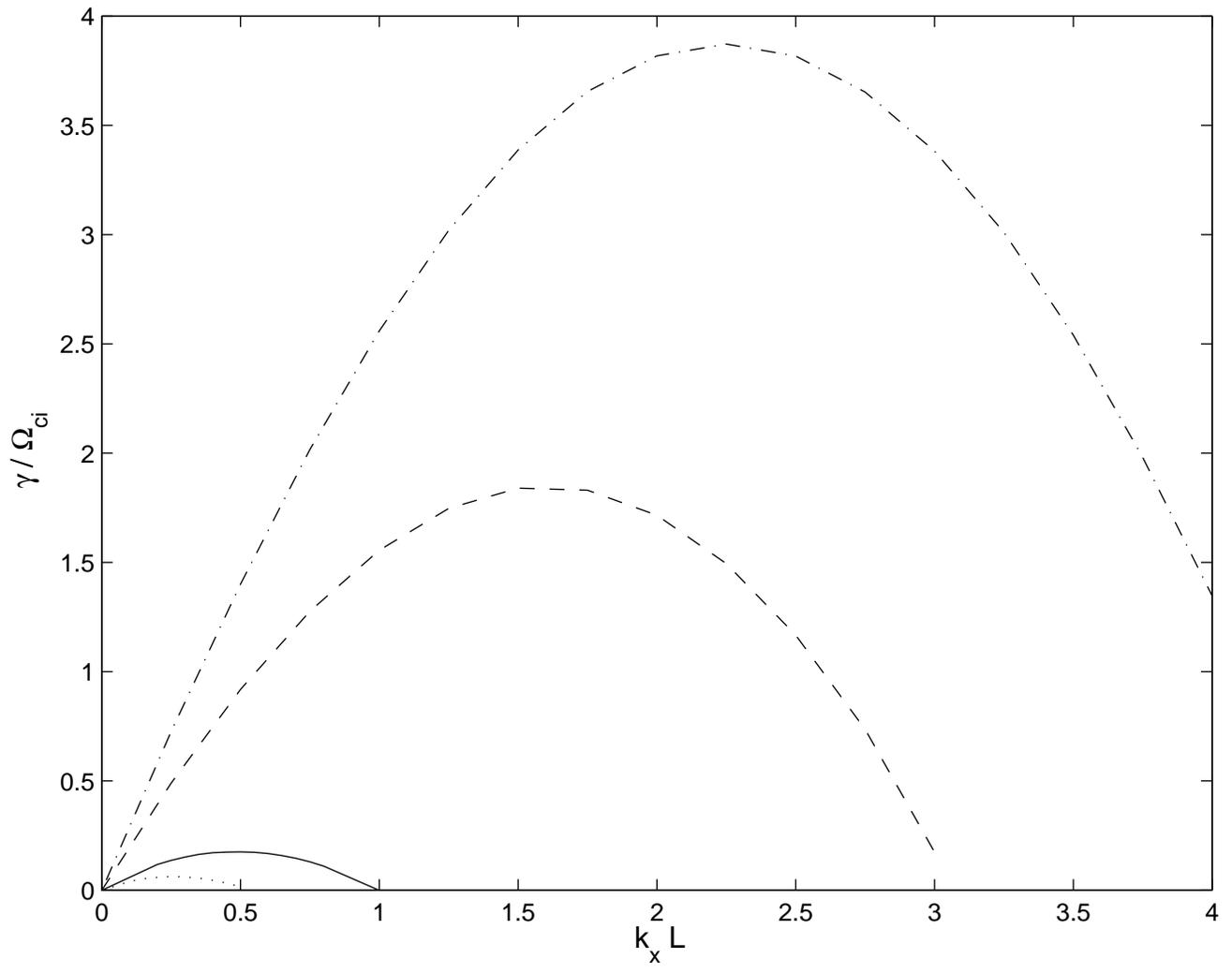
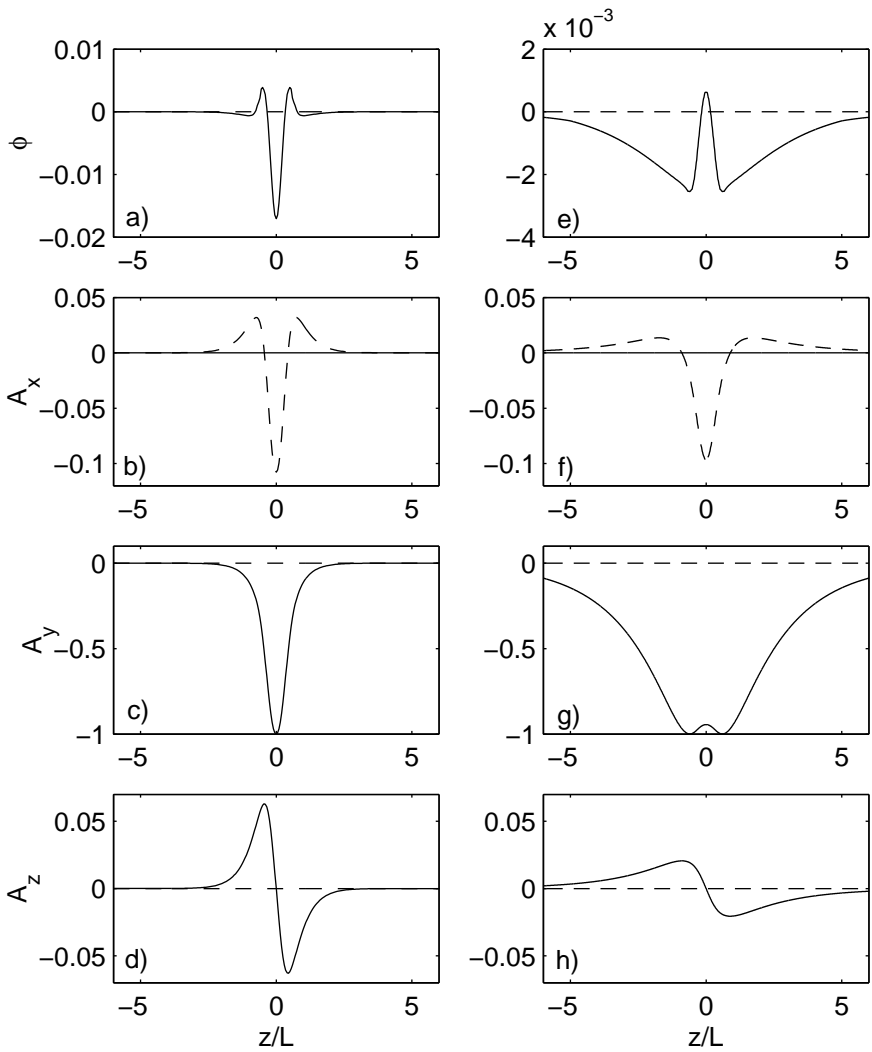
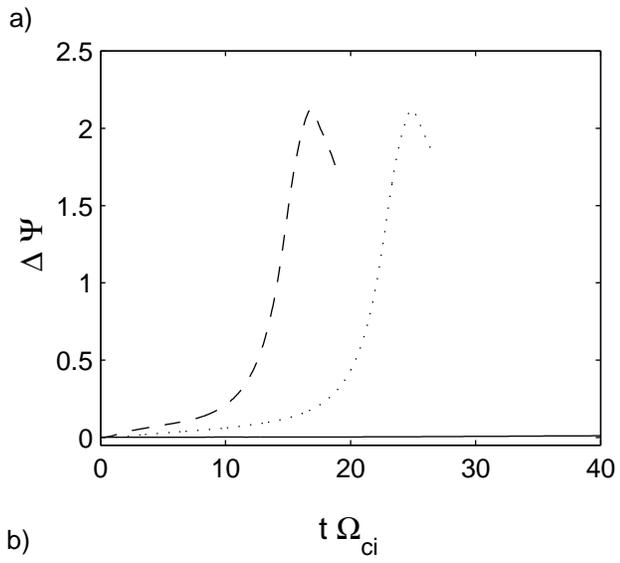
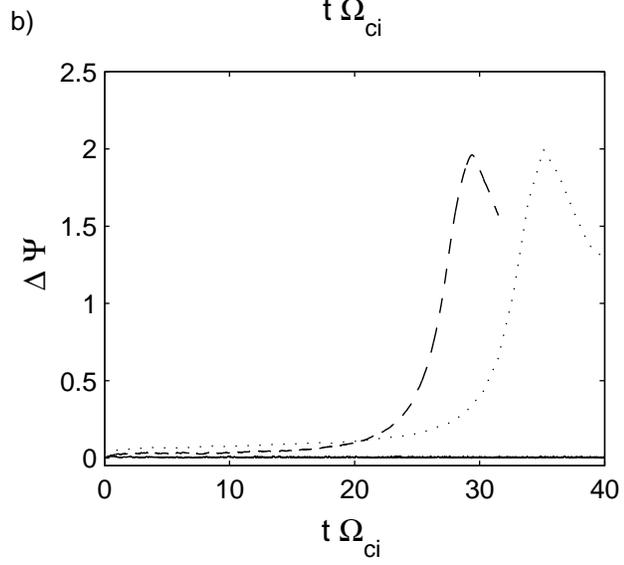
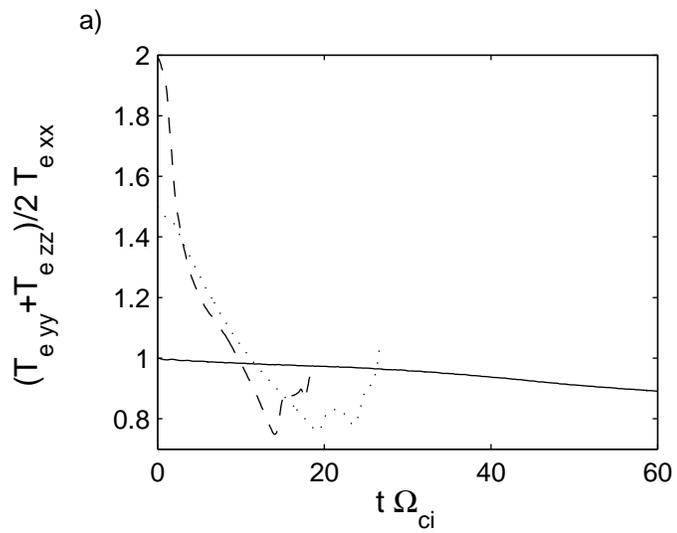
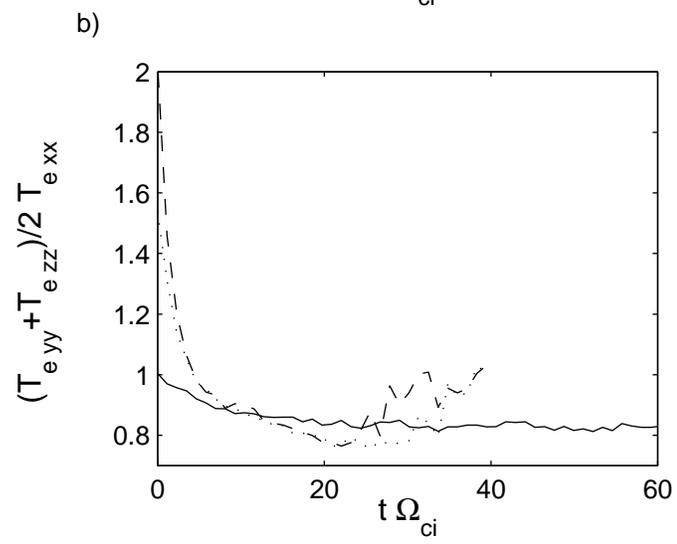
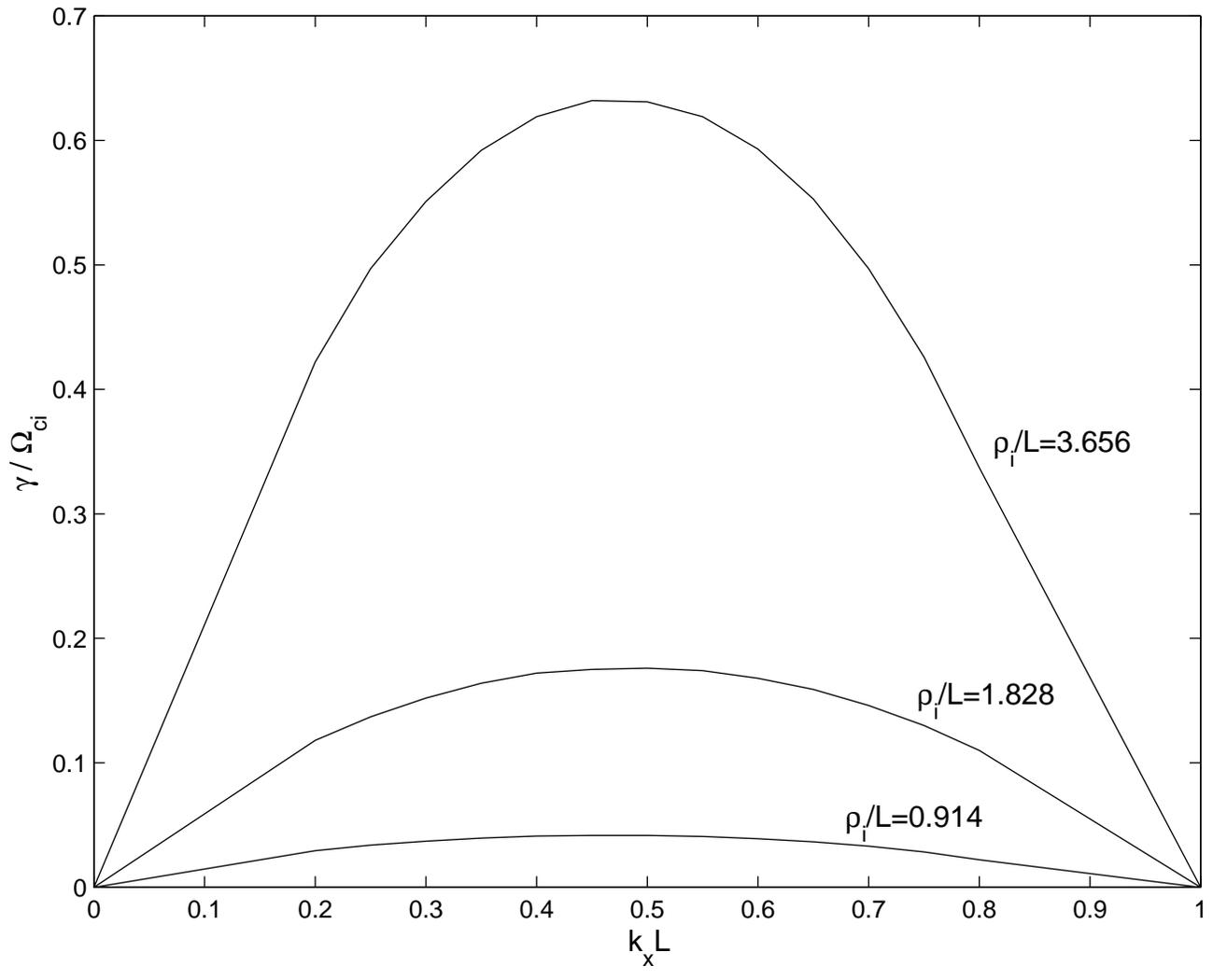
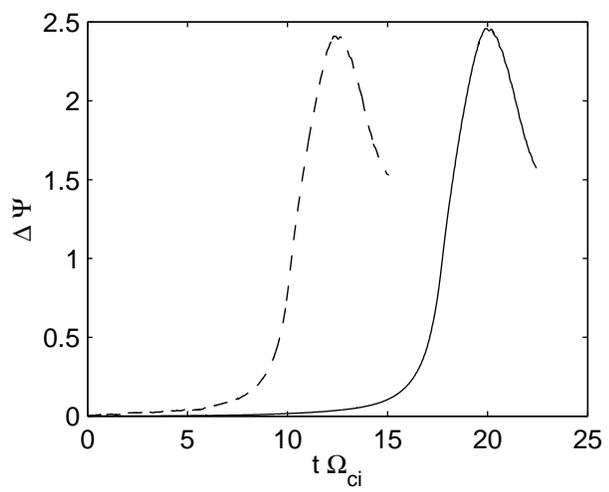
\end{document}